\documentclass[aps,prl,twocolumn,showpacs,superscriptaddress,showkeys]{revtex4-1}

\usepackage{amssymb}
\usepackage{amsmath}
\usepackage{amsfonts}
\usepackage{graphicx}
\usepackage{color}
\usepackage{xspace}
\usepackage{ulem}
\usepackage{mathtools}
\usepackage{hhline}

\usepackage{dcolumn}
\usepackage{bm}
\usepackage{hyperref}
\usepackage{comment}

\usepackage[hyperref,svgnames]{xcolor}

\newcommand{\gsim}{\lower.7ex\hbox{$\;\stackrel{\textstyle>}{\sim}\;$}}
\newcommand{\lsim}{\lower.7ex\hbox{$\;\stackrel{\textstyle<}{\sim}\;$}}

\newcommand{\eV}{{\, {\rm eV}}}
\newcommand{\neV}{{\, {\rm neV}}}

\newcommand{\MeV}{{\, {\rm MeV}}}
\newcommand{\GeV}{{\, {\rm GeV}}}
\newcommand{\TeV}{{\, {\rm TeV}}}

\newcommand{\kg}{{\, {\rm kg}}}
        
\newcommand*{\earth}{{\oplus}}

\def\beq{\begin{equation}}
\def\eeq{\end{equation}}
\def\bea{\begin{eqnarray}}
\def\eea{\end{eqnarray}}
\def\bitem{\begin{itemize}}
\def\eitem{\end{itemize}}
\newcommand{\bec}{\begin{center}}
\newcommand{\eec}{\end{center}}
\newcommand{\ba}{\begin{array}}
\newcommand{\ea}{\end{array}}

\def\inv{^{\raise.15ex\hbox{${\scriptscriptstyle -}$}\kern-.05em 1}}
\def\lbar{{\lower.35ex\hbox{$\mathchar'26$}\mkern-10mu\lambda}} 

\let\<=\langle
\let\>=\rangle

\let\+=\uparrow

\def\beqa{\begin{equation}\begin{aligned}}
\def\eeqa{\end{aligned}\end{equation}}

\newcommand{\AddrOxford}{Rudolf Peierls Centre for Theoretical Physics, University of Oxford, Oxford OX1 3PU, United Kingdom}
\newcommand{\AddrCoimbra}{Univ Coimbra, Faculdade de Ci\^encias e Tecnologia da Universidade de Coimbra and CFisUC, Rua Larga, 3004-516 Coimbra, Portugal}

\begin{document}

\title{Micro-Bose/Proca dark matter stars from black hole superradiance}

\author{John March-Russell} 
\affiliation{\AddrOxford}
\author{Jo\~{a}o G.~Rosa} \email{jgrosa@uc.pt}\affiliation{\AddrCoimbra}

\date{\today}

\begin{abstract}
We study the production of heavy, $\mu \gtrsim 1\TeV$, bosonic spin $s=0,1$ dark matter (DM) via the simultaneous processes of Hawking evaporation and superradiance (SR) from an initial population of small, $\lesssim 10^6\kg$, primordial black holes (PBHs).  Even for small initial PBH spins the SR process can produce extremely dense gravitationally-bound DM Bose or Proca soliton ``stars" of radius $\lesssim {\rm pm}$ and mass $\sim 10^{\rm few}\kg$ that can survive to today, well after PBH decay. These solitons can constitute a significant fraction of the DM density, rising to $\gtrsim 50\%$ in the vector DM case.
\end{abstract}


\maketitle



\section{Introduction}

The nature and production mechanism of the gravitationally inferred DM density is one of the most 
pressing physical mysteries.  Because we have no convincing evidence of the existence of {\it non}-gravitational interactions of DM it is vital to explore the physics of DM that interacts only or dominantly via gravity. This includes both possible production mechanisms and later phenomenology.
Here we explore a scenario where heavy DM is gravitationally produced by the decay of small PBHs, both through Hawking evaporation (HE) \cite{Hawking:1974rv,Hawking:1975vcx} as studied in \cite{Fujita:2014hha,Allahverdi:2017sks,Lennon:2017tqq,Hooper:2019gtx,Hooper:2020evu}, and via superradiance (SR) \cite{Arvanitaki:2009fg}. 
We focus on PBHs with initial mass $M_0\lesssim 10^6\kg$ that decay before Big Bang Nucleosynthesis, and that do not dominate the Universe's energy budget, so reheating occurs via, eg, inflaton decay.
SR production only occurs if DM is bosonic, and we consider both spin $s=0,1$ cases.
By `heavy' DM we mean of mass $\mu\gtrsim 1\TeV$, as distinct from the previously considered SR-production of ultra-light, $\lesssim 0.1\neV$, states such as the QCD axion or axion-like-particles  \cite{Arvanitaki:2009fg,Arvanitaki:2010sy,Pani:2012bp, Rosa:2012uz, Witek:2012tr, Brito:2014wla, Arvanitaki:2014wva, Arvanitaki:2016qwi, Baryakhtar:2017ngi, Rosa:2017ury, Baumann:2018vus, Cardoso:2018tly, Baumann:2019eav}.  

This change in particle mass leads to much different physics.
We find that from an initial quantum
fluctuation SR can now produce a sizeable fraction of the total DM density,
unlike the $\mu \lesssim 0.1\neV$ case. 
Moreover, strikingly,
SR can now lead to dense gravitationally bound DM soliton `micro-stars' that can survive to the present, long after the HE decay of the PBH itself. In brief, after the PBH is formed, SR initially dominates the BH evolution, producing a dense DM `cloud' surrounding the BH, extracting the BH's angular momentum $J$, so spinning it down and leading SR to become inefficient.  HE then dominates the BH mass loss, and part of the cloud is re-absorbed until the BH finally disappears. (We assume no ${\cal O}(M_P)$ BH remnant survives, though our conclusions would not change.)  The remaining cloud evolves to a `micro-Bose/Proca-star' (micro-BS/PS) of sub-atomic size and $\sim10^{\rm few}\kg$ mass.

Although not widely appreciated, an important fact is that SR can lead to macroscopic violation of an otherwise conserved global symmetry just like the complete HE of a BH formed from a global-charge-carrying state. The origin of this global symmetry violation is the same as for HE, namely the BH horizon boundary conditions, which in the SR case gives rise to the imaginary parts of the energies of the quasi-bound states.   Thus our SR-produced micro-stars can carry an apparently exact global quantum number as far as the non-quantum-gravitational part of their effective Lagrangian is concerned, and can have a lifetime, $\tau \gg 1/H_0$.  (It has not escaped our attention that this leads to new possibilities for baryogenesis, the details of which we study elsewhere.)  

As stated, minimally DM only interacts gravitationaly with the Standard Model (SM) in which case our scenario is essentially unconstrained despite the fact that, even for low PBH spins, a significant fraction of the DM may be in the form of micro-stars.
In a companion work we discuss the phenomena that these SR-produced DM 
micro-stars can give, depending on the details of any feeble non-gravitational interactions of DM with the SM.


\section{Black Hole Quantum Evolution}

We first briefly summarise the relevant features of SR. 
Field quanta of mass $\mu$ in a BH background have an effective gravitational coupling 
$\alpha\equiv {\mu M/M_P^2}$ where $M_P
\simeq 1.2\times10^{19}\GeV$
is the Planck mass (we set $\hbar=c=1$), 
and possess Hydrogen-like quasi-bound states described by quantum numbers $(n,j,\ell,m)$, with energies in the non-relativistic $\alpha\ll 1$ limit ${\rm Re}(\omega_n) \simeq \mu\left(1-{\alpha^2/ 2n^2}\right)$.    Here $n=\ell+1+n_r$, with $n_r$ the radial node number, and $j$ the total, $m$ the ${\hat z}$-axis, and $\ell=j$ (scalar case) or $\ell =j-1, j,j+1$ (vector) the orbital angular momentum quantum numbers.   These are quasi-bound states as the energies have an imaginary part which makes the bosonic field decay or grow.  Growth occurs for $\omega<m\Omega$ where $\Omega={\tilde a}M_P^2 /2(1+\sqrt{1-{\tilde a}^2})M$ is the Kerr angular velocity for a BH with dimensionless spin parameter ${\tilde a}\equiv J M_P^2/M^2$ ($0\leq {\tilde a}<1$, with ${\tilde a}=1$ extremal Kerr). 

The fastest growing modes have quantum numbers $n_r=0$, and $j=\ell=m=1$ for $s=0$, or $\ell=0$ and $j=m=1$ for $s=1$.  Although there may be many SR states, the dominant one will be populated exponentially faster than all others, so a good approximation is to take all DM particles to be produced in this state, the so-called `SR cloud'.  For $\alpha\ll 1$ and ${\tilde a}\ll 1$, the fastest SR growth rate $\Gamma_s \equiv 2{\rm Im}(\omega)$ (decay for $\Gamma_s<0$) is approximately
\begin{eqnarray}\label{SR_rate}
\Gamma_s(M,J)\simeq \left\{\begin{matrix}
\frac{1}{24}({\tilde a}-4\alpha)\alpha^8\mu~,\quad & s = 0\\
4({\tilde a}-4\alpha)\alpha^6\mu~,\quad & s = 1~.
\end{matrix}\right.
\end{eqnarray}
(Our figures use precise numerically-derived expressions for $\Gamma_s$ \cite{Cardoso:2018tly, Ferraz:2020zgi}. The SR regime is
$\alpha< {\tilde a}/4$ for ${\tilde a}\ll 1$ and $\alpha\lesssim 1/2$ for ${\tilde a}\sim 1$.) 
Note that, unlike HE, SR is more efficient in producing vectors not scalars. The number of bound particles in the SR cloud satisfies $dN/dt= \Gamma_s N$.

The PBH mass and spin will also be evolving due to the combined effects of HE and SR, thus implying an evolving $\Gamma_s(M,J)$ too.
The PBH dynamics is given by
\begin{eqnarray}\label{eq:evolneqns}
\!\!\!\frac{dM}{dt}&=& -e_T\frac{M_P^4}{M^2}-\mu\Gamma_s N,\nonumber\\
\frac{dJ}{dt}& = & -e_J\frac{J M_P^4}{M^3}-\Gamma_s N\,.
\end{eqnarray}
The energy-emissivity coefficient $e_T$ takes into account the contributions from all degrees of freedom (dof) with masses below the Hawking temperature $T_H=M_P^2/8\pi M$. Numerically, the spin-emissivity coefficient $e_J\simeq 7.8 e_T$ for a BH emitting all the SM dof at low ${\tilde a}$, with $e_J/e_T$ decreasing to $\simeq 2.4$ for ${\tilde a}$ close to extremality.  Note the
ratio can differ in SM extensions, eg, $e_J/e_T\simeq 5.2$ for the MSSM for ${\tilde a}\ll 1$. For our figures we use expressions for $e_T(\tilde{a})$ and $e_J(\tilde{a})$ inferred from the numerical data in \cite{Page:1976df,Page:1976ki,Chambers:1997ai,Taylor:1998dk} including all the SM states plus gravitons, since for $M\lesssim 10^6\kg$ one has $T_H\gtrsim 10\TeV$, with $e_T^{SM}\simeq 4.38\times10^{-3}$ at low $\tilde{a}$. The inclusion of the DM particle does not change this significantly, with e.g. $\Delta e_0=7.24\times 10^{-5}$ (spin 0) or $\Delta e_1=1.68\times 10^{-5}$ (spin 1) per dof at low PBH spin. 
For $e_T=e_T^{SM}$ a PBH of initial mass $M_0$ decays in time $t_{ev}=M_0^3 /3e_T^{SM} M_P^4$ when the radiation-era temperature $T_{ev}\simeq 1.4 ( 10.75/g_* )^{1/4} ( 10^6\kg/M_0 )^{3/2}\MeV$.  We conservatively take $M_0\lesssim 10^6$kg as an upper limit on the initial PBH mass given the stringent constraints on the abundance of PBHs that evaporate after BBN \cite{Carr:2020gox}.

\section{Superradiant production of DM}

SR is only efficient if it dominates the dynamics at early times, before HE damps the PBH angular momentum (as we later further discuss). In this case, DM is produced at the expense of the BH's rotational energy until it has spun down sufficiently so that $\Gamma_s=0$.
Each cloud particle carries angular momentum $\hbar$ and has energy $\simeq\mu$ in the $\alpha\ll 1$ limit.  To understand the number of DM particles produced by SR, note that changes in PBH $(M,J)$ are related by $\Delta M/M_0=\alpha_{0} {\tilde a}_0 (\Delta J/J_0)$, where the subscript `0' denotes the PBH initial state.  Thus SR has a more significant impact on $J$ than on $M$ and, to leading order, we can neglect the change in $M$ during a SR epoch, so $\alpha\simeq \alpha_{0}$ is quasi-constant.  As, for slowly rotating PBHs, the final spin parameter is ${\tilde a}\simeq 4\alpha$ one finds ${\Delta J/J_0}\simeq 1-{4\alpha/{\tilde a}_0}$.
(Our figures use precise numerical relations valid for large $\alpha,{\tilde a}$.)
From ${\Delta J/ J_0}$ the total number of DM particles produced by SR is:
\begin{eqnarray}\label{SR_number}
N_{SR}
\simeq  2.1\times 10^{27}{\tilde a}_0 \left(\frac{\Delta J}{J_0}\right) \left(\frac{M_0}{10^{6}\kg}\right)^2~.
\end{eqnarray}
This is the {\it maximum} number of DM particles produced by SR, since the PBH spin down due to HE will, in general, make SR less efficient \cite{foot1}. From Eqs.~(\ref{eq:evolneqns}), the evolution of the BH spin parameter is given by:
\begin{eqnarray}\label{eq:spinevol}
\frac{d\tilde a}{dt}={\tilde a}(-e_J+2e_T){M_P^4\over M^3}-\Gamma_s N{M_P^2\over M^2}(1-2{\tilde a}\alpha)~.
\end{eqnarray}
Firstly, from this we deduce that $\tilde{a}$ is sufficiently stable to build up a large SR-produced cloud before HE quenches SR if $\Gamma_s t_{ev}\gtrsim (e_J/e_T-2)\ln N_{SR}\sim \mathrm{few}\times 100$ in the PBH mass range of interest. This yields a $\tilde{a}$-dependent lower bound on the PBH mass.
Moreover, SR is not efficient at too small values of $\alpha$ and we find, in particular, that $N_{SR}$ can only be reached for $\tilde{a}_0\gtrsim 0.01$.

Secondly, even in such parametric regions, the PBH continues to spin down after $N_{SR}$ is attained and the SR condition is saturated
(at $\tilde{a}\simeq 4\alpha$ for small PBH spin). This implies that the DM cloud will enter a regime where $\Gamma_s<0$ and it is at least {\it partially reabsorbed}. This reabsorption 
can balance the effect of HE on ${\tilde a}$ (though not $M$),
at least until HE becomes too fast at the end of PBH decay.
So there is a finite-duration quasi-stationary regime, $d{\tilde a}/dt\simeq 0$, close to the SR threshold ${\tilde a}\lesssim 4\alpha$, such that $\Gamma_s\simeq 4\mu(-e_J+2e_T)/N$, as shown in Fig.~1.

\begin{figure}[h!]
\centering\includegraphics[scale=0.134]{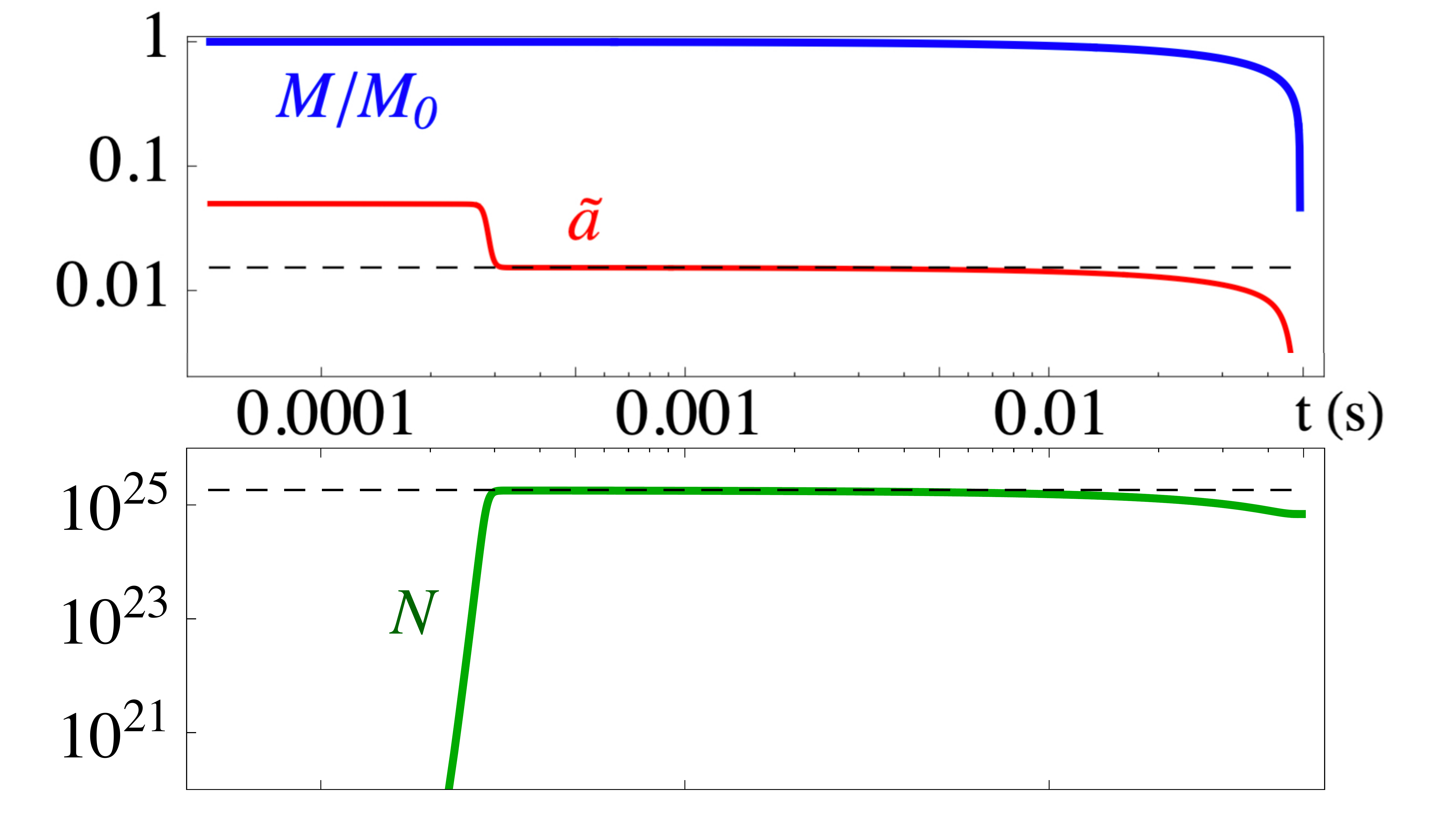}\vspace{-0.2cm}
\caption{Evolution of BH mass and spin (top) and scalar DM number $N$ in the SR cloud (bottom) for $M_0\!=\!5\times10^5$kg, $\tilde{a}_0\!=\!0.05$, $\mu\!=\!2$TeV ($\alpha_0\simeq0.004$). Note (1) in this low BH spin case SR extracts a large fraction of the BH's spin but not its mass; (2) BH/cloud enter a quasi-stationary regime with $N, \tilde{a}$ 
near the SR threshold (dashed lines); and (3) the cloud is then partially reabsorbed (here by factor $\epsilon_{SR}\simeq0.38$) as HE spins the BH down.}
\label{dynamics}
\end{figure}

Concerning the cloud reabsoption, for $N\simeq N_{SR}$ and 
before the BH loses a significant mass fraction, we have $|\Gamma_s| t_{ev}\simeq {4(e_J/e_T-2)\alpha/3({\tilde a}_0-4\alpha)}$. If this quantity is small the number of cloud particle remains close to $N_{SR}$ until the BH starts losing a significant mass fraction. At this stage the consequent decrease in the coupling $\alpha$ quenches $|\Gamma_s|$ and reabsorption effectively stops. The estimated overall cloud decrease ${\rm exp}(-|\Gamma_{s}| t_{ev})$ is moderate if $\alpha \lesssim  3{\tilde a}_0/4(e_J/e_T-2)$, ie, away from the SR threshold. 
However, for larger values of $\alpha$ the number of DM particles decreases significantly below $N_{SR}$ before $t_{ev}$, which in turn increases the absorption rate, leading to a runaway effect where the system exits the quasi-stationary regime and both $N$ and $\tilde{a}$ quickly decrease. In such a regime the DM cloud is transient, being fully reabsorbed by the PBH.
Both the inefficiency of SR at low $\alpha$ and the significant reabsorption due to HE at large $\alpha$ can be encoded in a  SR efficiency parameter $\epsilon_{SR}<1$ which we numerically compute in our figures. 

To determine the fraction of the total DM yield from SR, we compare $\epsilon_{SR}N_{SR}$ with the number of DM particles produced by HE. Following \cite{Lennon:2017tqq}, this is well approximated by $N_{HE}\simeq (f_s/2e_T)\left({M_0/ M_P}\right)^2$.
Here $f_s$ is a `grey-body number factor', and for both this and $e_T$ we may use the low-$\tilde{a}$ values, with $f_0\simeq 6.66\times 10^{-4}$ (spin 0) or $f_1 \simeq 7.4\times10^{-5}$ (spin 1) per dof, since the bulk of the HE-emission of particles occurs when the BH has already lost most of its spin due to both HE and SR. (For $\mu \lesssim 3.2T_{H,0}$, the number of HE produced DM particles decreases by $(3.2T_{H,0}/\mu)^2$. For simplicity we will omit this factor in the presented formulae).  
Assuming a massive complex scalar field with 2 dof, or a Proca vector field with Stuckelberg mass and 3 dof, the total number of DM particles generated by HE per PBH is
\begin{eqnarray} \label{evaporation_number}
N_{HE}\simeq \left\{\begin{matrix}
{3.2}\times 10^{26}\left({M_0\over10^{6}\kg}\right)^2~, \quad s=0\\
{5.3}\times 10^{25}\left({M_0\over10^{6}\kg}\right)^2~, \quad s=1
\end{matrix}\right.~.
\end{eqnarray}

The ratio between the DM SR and HE yields is thus:
\begin{eqnarray} \label{Y_ratio}
{Y_{SR}\over Y_{HE}}= \epsilon_{SR}{N_{SR}\over N_{HE}}
\simeq  \epsilon_{SR}{\Delta J\over J_0}\times \left\{\begin{matrix}
{{\tilde a}_0\over 0.15} ~, \quad s=0\\
{{\tilde a}_0\over 0.025} ~, \quad s=1
\end{matrix}\right.~.
\end{eqnarray}
As $\Delta J/J_0\sim 0.1$-1 in the parametric range of $\alpha$ for which $\epsilon_{SR}\sim {\cal O}(1)$, SR can thus account from a few percent to the majority of DM if PBHs have natal spins ${\tilde a}_0\gtrsim 0.03$.  Note that PBHs formed through the collapse of $\mathcal{O}(1)$ density fluctuations gain spins ${\tilde a}_0\sim 10^{-2}$ once they re-enter the Hubble horizon in the radiation-dominated era \cite{Chiba:2017rvs, Mirbabayi:2019uph, DeLuca:2019buf}. (The radiation-era horizon mass is $M_H \sim 10^{23} g_*^{-1/2} (\TeV/T)^2 \kg$, so the PBHs of interest could have formed very early.)   Studies \cite{Harada:2017fjm, Arbey:2021ysg} also suggest that during an early matter era (eg, from late decay of the inflaton field) PBHs can have large natal spins, even near extremal, as a result of anisotropic collapse in a pressureless environment.

\begin{figure}[h!]
\centering\includegraphics[scale=0.42]{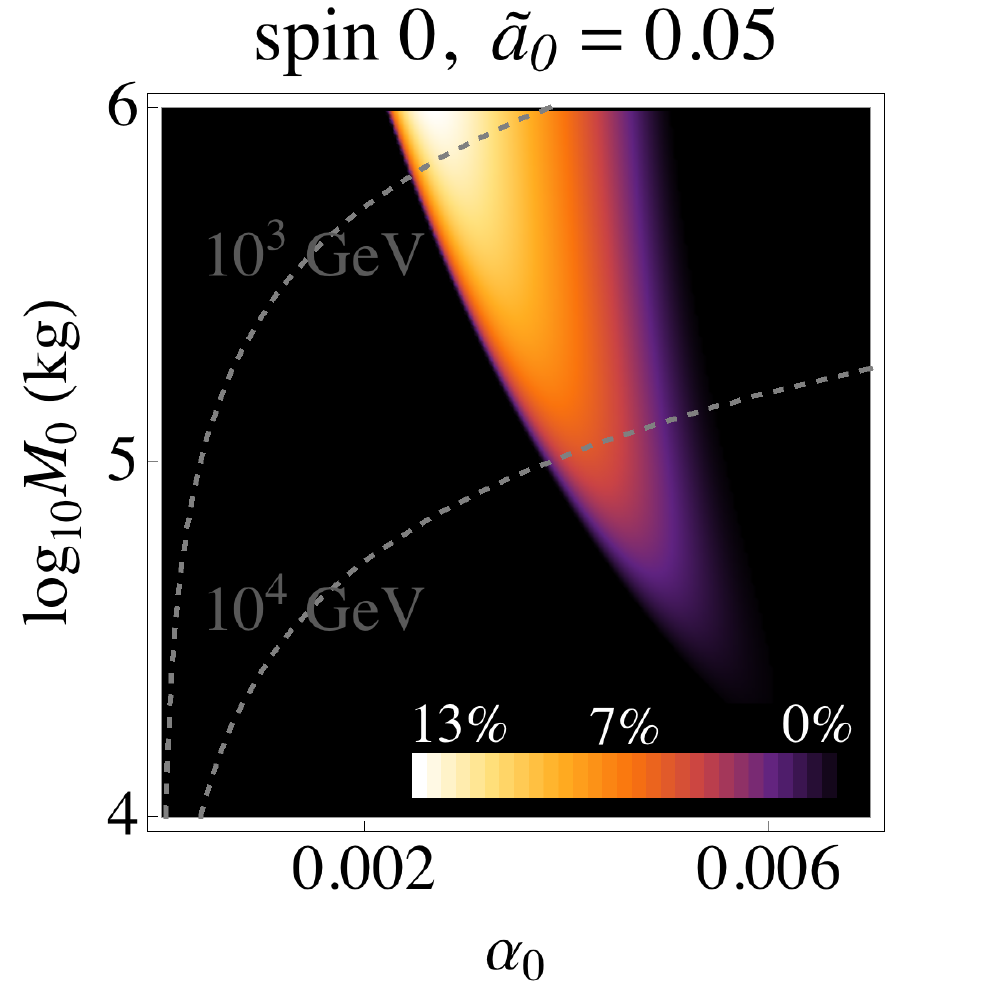}
\centering\includegraphics[scale=0.42]{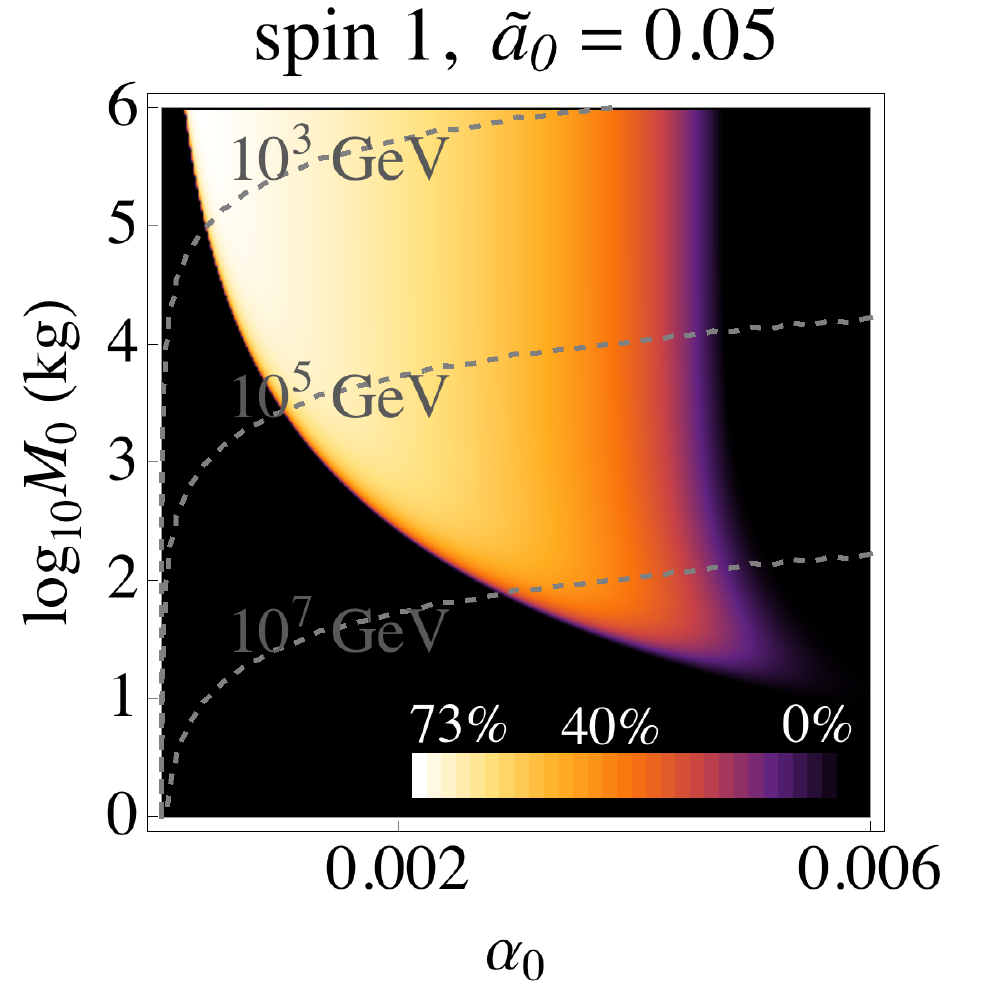}
\centering\includegraphics[scale=0.42]{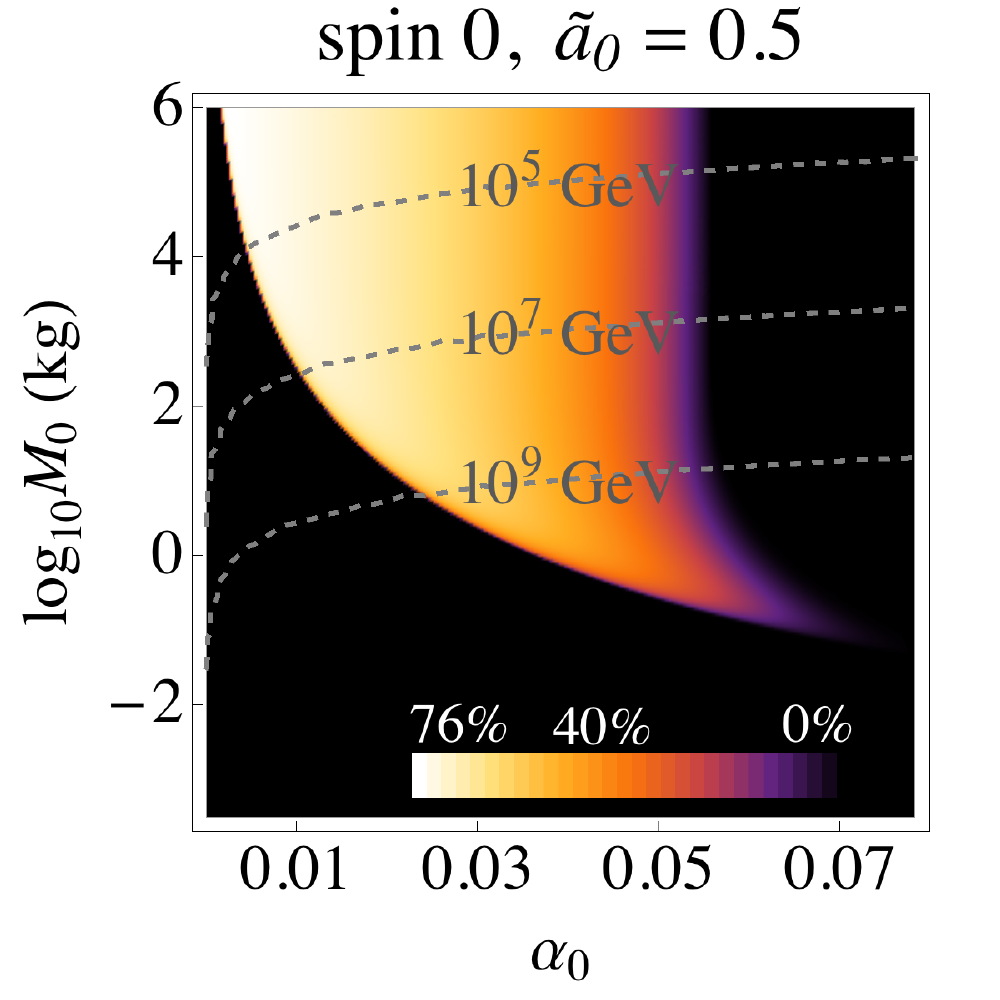}
\centering\includegraphics[scale=0.42]{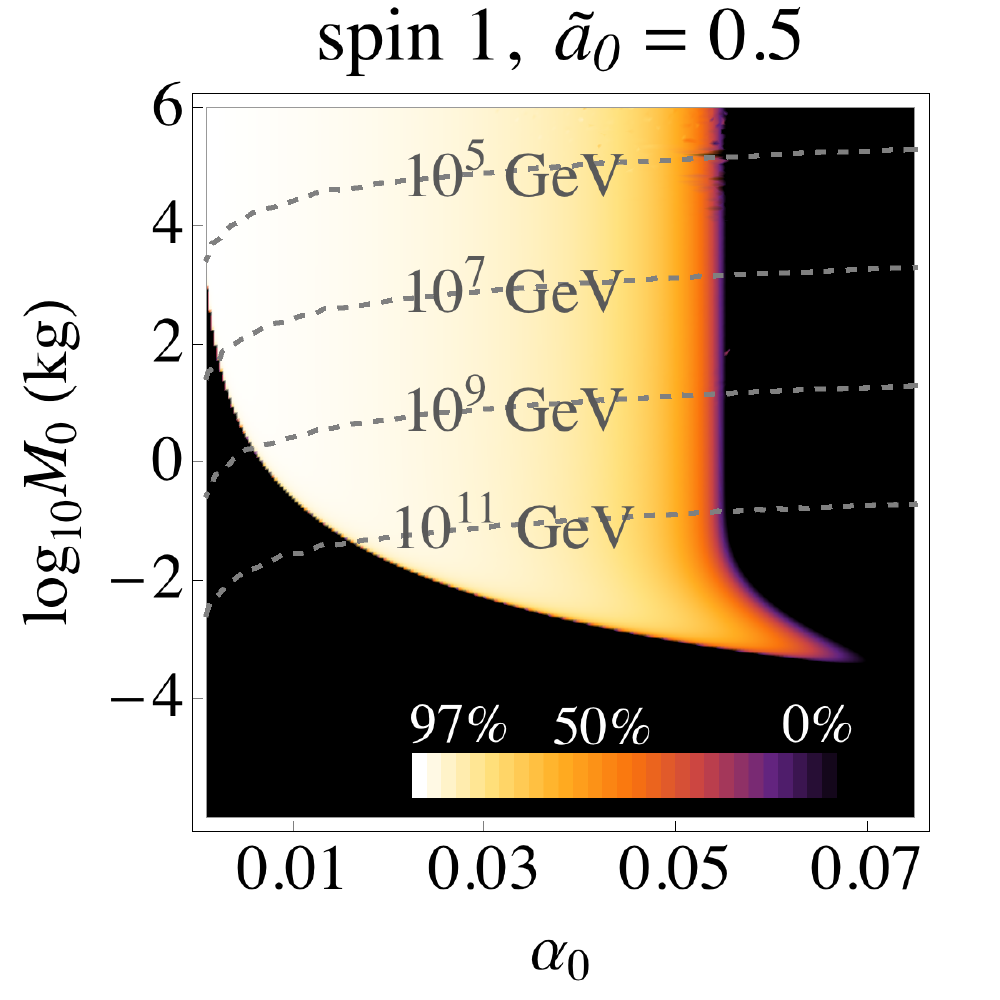}
\caption{Fraction of the DM yield produced through SR for spin 0/spin 1 DM particles (left/right), assuming a low/high PBH spin (top/bottom), as a function of the initial PBH mass $M_0$ and DM-mass-dependent coupling $\alpha$. Outside the shaded region SR is either forbidden or inefficient compared to HE.  Dashed gray curves are iso-DM-mass-contours as labelled.}
\label{DM_fraction}
\end{figure}
The resulting final SR cloud DM fraction is shown in Fig.~2, where we have computed the SR efficiency $\epsilon_{SR}$ by numerically solving Eqs.~(\ref{eq:evolneqns}) for different values of $M_0, \tilde{a}_0$ and $\alpha_0$. 
For both spin-0 and spin-1 DM, as well as for both small and large PBH spins, there is a finite region in the ($\alpha_0$, $M_0$) plane where SR DM production is efficient, this region being broader for vector DM, for which SR/HE is more/less efficient than in the scalar case, as well as for larger PBH spins. In all cases, the DM mass $\mu\gtrsim1 $ TeV and may even take values close to the GUT scale, showing that SR can be an efficient way of gravitationally producing heavy DM particles.

Turning to the total DM yield, we have $Y_{DM} \equiv {n_{DM}/ s} = (N_{SR}+N_{HE}) {n_{BH}/ s }$ where
$s$ is the entropy density and $N_{SR}$ and $N_{HE}$ are given in Eqs.(\ref{SR_number},\ref{evaporation_number}).  Note that $Y_{DM}$ depends on the initial PBH density at formation,  $n_{BH}^0$, unlike the
`slow' case considered in \cite{Lennon:2017tqq} where the PBHs both dominate the energy density and themselves
give rise to SM reheating via HE.  Reproducing the observed DM density requires $\mu Y_{ev}\simeq 0.43\eV$ which fixes $n_{BH}^0$.
Similarly to previous discussions of DM production by PBH HE 
\cite{Fujita:2014hha,Allahverdi:2017sks,Lennon:2017tqq} this is straightforward to achieve consistently with all other constraints (eg, dark radiation from graviton emission \cite{Hooper:2020evu}).


\section{Clouds and micro-Bose/Proca stars}

Since a significant cold DM fraction can arise from SR clouds we should investigate their fate.  During HE of the central PBH, a second effect of the HE, in addition to partial reabsorption of the cloud particles, is that initially the cloud's radius increases with the gravitational `Bohr radius', $r_B=1/\alpha\mu$, as $\alpha \simeq M(t)$ in the regime $M(t) \gg M_c$ where $M_c$ is the cloud's mass.   In fact,  as long as the PBH evaporates slowly, the DM particles will remain bound in the same state dominantly populated by SR, the only difference being the increasing cloud radius.  This is exactly analogous to an adiabatic change in the Hamiltonian of a quantum system, since the quasi-bound states are described by a Schr\"odinger equation.  

Adiabaticity here refers to the change in the gravitational binding energy,
$\omega_B \simeq \alpha^2\mu/2n^2$ (in the regime $M(t) \gg M_c$), so we may define an adiabaticity parameter
$\xi \equiv {\dot\omega_B/\omega_B^2}= {4n^2e_T M_P^2/ \alpha^{3}M^2}$.
Since $\xi\sim M(t)^{-5}$ for $M(t) \gg M_c$, at the final stage the HE process can possibly become non-adiabatic, corresponding effectively to a sudden disappearance of the PBH from the cloud's dynamics point of view.  As in a sudden change in a quantum Hamiltonian, the cloud will transition into a superposition of the eigenstates of the new Hamiltonian, both bound and free. In the absence of the BH (and DM self-interactions, a topic considered in 
the companion paper), the only potential is the cloud's self-gravity, so the DM particles can either become free with the cloud dispersing away as a collection of cold DM particles or remain in a self-gravitating quantum soliton configuration. 

The most interesting scenario is one where the majority of the cloud remains self-bound after the PBH decays.  Although full evaluation of this possibility requires numerical simulation, since the cloud's dynamics becomes non-linear, we may assess how likely it is by comparing $M_c$ to the PBH's mass, $M_*$, when $\xi\simeq 1$, i.e. possibly non-adiabatic.  If $M_*\lesssim M_c$, the bound-state wavefunctions before and after the sudden HE of the PBH should have a large overlap, a significant fraction of the cloud particles remaining self-bound. 
We find that for spin-0 
\begin{eqnarray} \label{adiabaticity_ratio}
{M_*\over M_c}
\!\simeq\! {{0.44\over \epsilon_{SR}}}\left({\Delta J_0\over J_0}\right)^{\!\!-1}\!\!\!\left( {0.05\over {\tilde a}_0} \right)\!\!\left({0.003\over \alpha}\right)^{\!\!8/5}\!\!\left({10^6\kg\over M_0}\right)^{\!\!2/5}\!\!\!\!\!\! ,
\end{eqnarray}
so $M_*/M_c \lsim 1$ is possible even for the smallest values of $\alpha$ and ${\tilde a}_0$, provided that cloud reabsorption is not very significant, i.e. in the parameter range for which the SR fraction is larger. A similar result holds in $s=1$ case. We thus expect much of the SR cloud to survive the decay of the central PBH in a considerable part of the parameter space, and particularly for heavier PBHs.  So an exciting possibility is that a non-negligible part of the DM density is bound in these self-gravitating clouds with a large number, $N\sim 10^{25}$, of particles.

As mentioned in the introduction self-gravitating soliton objects are known as (scalar) boson (or Bose) stars (BS) or Proca stars (PS) for $s=0,1$ fields, respectively.  The scalar rotating solutions with $J=L\neq 0$ have a toroidal shape similar to the SR clouds with the same quantum numbers, but are unstable against non-axisymmetric perturbations \cite{Sanchis-Gual:2019ljs, DiGiovanni:2020ror, Dmitriev:2021utv}, breaking up into one or more of the more stable non-rotating scalar BS solitons, which are spherical.  PS with $J\neq 0$ but $L=0$ have a spheroidal topology and numerically appear to be stable.  So, for both $s=0,1$ DM, the end-state of the SR clouds after PBH decay is expected to be either a single or possibly a number of spheroidal self-gravitating solitons.  

These solitons are described by the non-relativistic Ruffini and Bonazzola \cite{Ruffini:1969qy} solutions of radius $R_{BS}\simeq {10M_P^2/\mu^2M_n}$, and are hence of sub-atomic size: 
\begin{eqnarray}\label{nugget_radius}
\!\!\!\!\!\!R_{BS}\!\simeq\!\! {5.5\over \epsilon_{SR}f_{BS}}\!\left(\!{0.003\over \alpha_0}\!\right)^{\!\!3}\!\! \left({0.05\over {\tilde a}_0}\right)\!\! \left(\!{M_0\over 10^6\kg}\!\right)\!\!\! \left({\Delta J_0\over J_0}\right)^{\!\!-1}\!\!\!\mathrm{pm}
\end{eqnarray}
using the estimate for the boson star ($s=0$) mass 
\begin{equation}\label{nugget_mass}
\!\!\!M_{BS}\! \simeq \! 150\epsilon_{SR}f_{BS}\! \left({\alpha_0\over 0.003}\right)\!\! \left(\!{{\tilde a}_0\over 0.05}\!\right)\!\! \!\left(\!{M_0\over 10^6\kg}\!\right)\! \!\left(\!{\Delta J\over J_0}\!\right) \mathrm{kg}.
\end{equation}
Here $f_{BS}\lesssim 1$ parameterizes the loss of DM cloud particles in the transition from SR cloud to micro-BS/PS and, for $s=0$, also in the decay of the rotating toroidal solitons \cite{Sanchis-Gual:2019ljs}.
The SR process implies that in the vector DM case the resulting PS is maximally spin-polarized with spin $|J_{PS}|=N\simeq M_{PS}/\mu$. The physics of these solitons is quite rich as investigated in \cite{Visinelli:2021uve,Adshead:2021kvl,Jain:2021pnk,Gorghetto:2022sue}
though here we are concerned with a different production mechanism and
typically a much smaller soliton mass.  

Since $\TeV \ll M_{BS} \ll M_\earth$ the dark BS/PS are both rare and hard to detect if they have no non-gravitational interactions with the SM.  From the inferred local density of DM in the solar neighborhood their approximate flux is $\Phi \simeq 5\times10^{-4} f_{SR}f_{BS}\left(10 \kg/M_{BS}\right)\ \mathrm{km^{-2}yr^{-1}}$, where $f_{SR}$ is the fraction of the DM yield produced via SR shown in Fig.~2. 
Signatures will largely depend on the nature of non-gravitational interactions between DM and SM particles, as we will explore in a future companion paper. Generically, the latter can be small enough not to affect the dynamics of SR DM production, rendering the detection of individual DM particles virtually impossible, but nevertheless lead to observable signals of dark BS/PS due to the large number of DM particles contained in each of these microscopic solitons. In particular, we may expect large coherent enhancements of interaction cross-sections with SM nuclei \cite{Hardy:2015boa}.

\vspace{0.2cm}

{\bf Acknowledgments:} We thank Nicolas Sanchis-Gual for helpful discussions. This work was supported by the FCT research grant no.~CERN/FIS-PAR/0027/2021. JGR is also supported by the CFisUC project No.~UID/FIS/04564/2020.


\begin{thebibliography}{99}
\bibitem{foot1}
For BSM scenarios with many light spin-0 states this is not necessarily so \cite{Calza:2021czr}.  For simplicity we ignore this exception, so during the epoch where HE is the dominant process the PBH has small ${\tilde a}$.
\end{thebibliography}

\end{document}